\documentclass[12pt]{article} \textheight 23cm
\usepackage{graphicx}
\usepackage{amsmath}
\usepackage{color}

\usepackage{amsfonts}
\usepackage{float}
\usepackage{setspace}

\begin{document} 
\begin{center} 
{\large Memory of rheological stress in polymers using Fractional Calculus 
} \vskip 0.2cm
 Archishna Bhattacharyya$^1$, Pratyusha Nandi$^1$, Somasri Hazra$^2$ and Tapati Dutta$^{1,2,*}$\\
\vskip 0.5cm
$^1$  Physics Department, St. Xavier's College, Kolkata 700016, India\\
$^2$ Condensed Matter Physics Research Centre, Physics Department, Jadavpur University, Kolkata 700032, India\\
$^*$ Corresponding author: email:tapati$\textunderscore$mithu@yahoo.com,tapati$\textunderscore$dutta@sxccal.edu\\
 Phone: 919903872208 
\end{center}
\vskip 1cm
\noindent {\bf Abstract}\\ 
The rheological properties of viscoelastic materials like polymer melts are greatly affected by factors like salinity, temperature, concentration and pH of the solution. In this study, the `memory of the stress' affected by each of these factors is shown to be trapped in the \textit{order of the fractional derivative} of the dynamical equation describing stress and strain in the material. To demonstrate this, the rheological properties of the polymer melt hydrolysed polyacrylamide (HPAM) have been modelled using a two-element Maxwell model. The model has successfully reproduced existing  experimental data on elastic modulus and complex viscosity for these `stress' factors, besides predicting the development of creep compliance with shear rate. The work also establishes that it is possible to `tailor' a particular rheological property by suitably tuning a pair of properties, \textit{complementary conjugates}, that offset each other's effects on the rheology. The study shows that HPAM has at least two pairs of \textit{complementary conjugates} in (i)temperature and pH, and (ii)concentration and pH. Further it is shown that the variation of viscosity with shear rate
shows a power law behaviour for almost all variations in stress parameters. Our modelling using fractional calculus establishes that the fractional order derivative $q$ which is recognised as a
memory index to emergent phenomena, shows an inverse relationship with respect
to the power-law exponent $\alpha$; the higher the memory index $q$, the smaller is the
power-law exponent $\alpha$.


 \pagestyle{plain}
 \section*{Introduction} \label{intro}
  
 Polymers have extensive industrial applications as adhesives for petroleum pipeline coating and the packaging industry, as a crude oil pour-point depressant,$^{1,2}$ displacement agents in oil recovery ,$^{3}$ oil well cement modifier,$^{4}$ as rheology control additives in consumer products such as shampoos, creams, and in clinical medicine etc. Hence the study of their rheology becomes pivotal amongst complex fluids in general. The most important application of hydrolysed polyacrylamide (HPAM), which motivates this study, is in Enhanced Oil Recovery (EOR), a technique employed in the tertiary stage of oil recovery.$^{5,3}$ 
 The generic term 'polymeric liquids' includes the entire spectrum of possibilities ranging from dilute polymer solutions, through concentrated solution regime to polymer melts.$^{6}$ The rheological properties observed in all of the above forms is attributed to long chain molecules and the length of the chain is the most influential property in determining their rheology among others. Another important phenomenon relevant to our study is entanglement. When the chains are long enough, the intermolecular association occurring is known as entanglement which gives rise to high elasticity effects such as high extensional viscosity, irrespective of whether the polymer is in molten state or in non- dilute solutions.$^{6,7}$  In general, the non-Newtonian behaviour of polymers is manifest in the rate of loading. For a certain large strain, the stress tends to attain its final value at different paces, depending on how gradually the load is distributed to the system. They also display creep, i.e., an increasing deformation under sustained load, the
rate of strain depending on the stress. At rest, the chains are randomly entangled, the entanglement increases with increasing polymer concentration in the solution, but they do not set up a definite geometric structure because the electrostatic forces present are predominantly repulsive. HPAM is an anionic polyelectrolyte with carboxyl groups present at the ends of its chains. It's molecular weight is 2.3 $\times(10^7)$ g/mol. When the fluid is in motion, the chains tend to align themselves parallel to the direction of flow, this tendency increases with increasing shear rate and is the reason for decrease in effective viscosity.
The chief physical property distinguishing a polymer from other complex fluids is the linearity of its chain. In general polymers are characterised by high molecular weights $(1000 - 10^9)$ g/mol and high viscosities. A phenomenon exhibited by polymer melts due to long linear chains is \emph{reptation.} It is the thermal motion of very long, linear, entangled macromolecules in polymer melts.

One of the simplest and most important characters of polymeric liquids is the existence
of an observable microscopic time scale. For regular liquids the timescales
of molecular motion are in the order of $10^{-15}$ seconds, associated with molecular
translation. In polymeric liquids, apart from this small time scale, there is an important
timescale associated with large scale motions of the polymer chain itself, in the
liquid they are suspended in (solutions or melts). This could be from microseconds
to minutes. Since many observable and processing time scales are of similar
order, the ratio of these two time scales becomes important. The large scale microscopic
motions are usually associated with the elastic character of the polymeric liquids. The
relaxation time $\lambda$, is the time associated with large scale motion (or changes) in the
structure of the polymer. The macroscopic time scales arise from two sources: (i)strain rate, measured by
the local shear rate $\dot{\gamma}$ for shearing flows, or the local elongation rate $\epsilon$ for extensional
flows; (ii) a dynamic timescale $t_d$ associated with the motion of the fluid
packets themselves. Besides material properties, characterising a material as a solid or fluid depends on the time scale of observation.  Glass on a window pane appears solid for all practical purposes, but on a time scale of centuries it flows.
In most practical applications the fluid packets undergo a non-uniform stretch history.
This means that they could have been subjected to various strain rates at various
times in their motion. Therefore no unique strain rate can be associated with the
flow. In these cases it is customary to refer to the Deborah number defined as the
ratio of the polymeric time scale,$^{8,9}$ $\lambda$ to their dynamic or flow timescale $t_d$
$$De = \lambda/t_d$$
For $De \ll 1$ the polymer relaxes much faster than the fluid packet traverses
a characteristic distance, and so the fluid packet is said to have `no memory of
its state' a few $t_d$ back. On the other hand for higher $De$, the polymer is said to have not
sufficiently relaxed and the state a few $t_d$ back can influence the current motion of the packet.

 The use of fractional derivatives in defining complex dynamics is a powerful tool in various disciplines,$^{10-15}$  with the fractional order being identified as an index of `memory' where emergent phenomena is influenced by history of stress induced in the system.$^{16-20}$  Studies on rheological modelling of biological fluids ,$^{21,22}$ natural rubber and polymers,$^{23-25}$ polymeric glasses and gelling systems are some of the areas of research interest.$^{26}$ Fractional differential equations have been used to describe the non-linear viscoelastic behaviour of rheology in fluids using fractional derivatives.$^{27,28}$ The integral equation that defines fractional derivative, either Riemann-Liouville (RL) or Caputo, are non-local in definition. As the calculation requires knowledge of all previous states of the function from the start, it therefore incorporates history. The fractional definition is also a convolution operation of a kernel function which is a decaying power law with function or derivative of function. This gives relaxation phenomena with memory and thus history gets imbibed in the RL or Caputo definitions. The classical Caputo or RL definitions have singular decaying power law memory, that gives non-local character. Recent advances use non-singular decaying kernel functions in the convolution definition, as in Caputo-Fabrizio and Atangana-Baleanu fractional derivatives etc. which too are non-local, and thus have history inbuilt. However the fractional derivative formed by ‘conformable-fractional derivative’ is local in nature, and thus will not emulate memory or history or heredity. Thus, unlike integer derivative calculus that depends on local information,  fractional order derivatives depend not only on local conditions of the evaluated time but also
on the entire history of the function. This fact is often
useful when the system has a long-term `memory' and
any evaluation point depends on the past values of the
function. 

 Through this study we present: (i) an appropriate model that captures the viscoelastic nature of the polymer melt hydrolysed polyacrylamide (HPAM) under different stress conditions like variation in salinity, temperature, concentration and pH; (ii)observe that the rheology of HPAM is extremely sensitive and can change by orders of magnitude for slight changes in some of these parameters; (iii)verify that only the use of fractional order derivatives can appropriately reproduce experimental data of the polymer melt; (iv)claim that it is possible to `tailor' complex viscosity of HPAM by appropriately tuning certain parameters that are complementary in their effects (v)establish that the fractional order of the derivative does indeed trap the `memory' of stress history  that the polymer melt had undergone to achieve a particular rheological property. (vi)demonstrate that the order of the fractional derivative $q$ is inversely related to the exponent $\alpha$ of the power law behaviour between viscosity and shear rate, for each of the stress variables. 
Viscoelastic systems, especially polymers have been modelled earlier using fractional derivatives as found in.$^{27,28}$ Recently,$^{29}$  a more evolved model using two spring-pots i.e. 4 parameters, was used by Jaishamkar et al. to characterise polymeric system. However we have economized on free parameters by employing a 3-parameter Maxwell model with a simple spring and a modified dashpot (spring pot) in series, to match available experimental data on HPAM successfully. , to predict properties of the HPAM melt. We have further predicted properties of the melt emerging from our test model.

 \section*{Materials and Methods} \label{m_and_m}
The polymer melt that has been been studied is hydrolysed polyacrylamide (HPAM).

In neutral water, HPAM behaves as an
anionic polyelectrolyte, with a high ionic strength. The electrostatic
interaction between carboxyl groups in HPAM molecular chains makes the molecular
coil extended. While a higher hydrolysis degree can cause an increase in viscosity,
it also increases its sensitivity to
salt. This can in turn lead to flocculation that decreases the viscosity of the solution. This is responsible for the  complex
rheology of HPAM. The experimental data of rheology of HPAM have been obtained from the works of Zhang et al.,$^{30,31}$ and Choi et al..$^{32}$

\subsection*{Viscoelastic modelling}

\begin{figure}[H]
	\begin{center}
	\includegraphics[width=15.2cm]{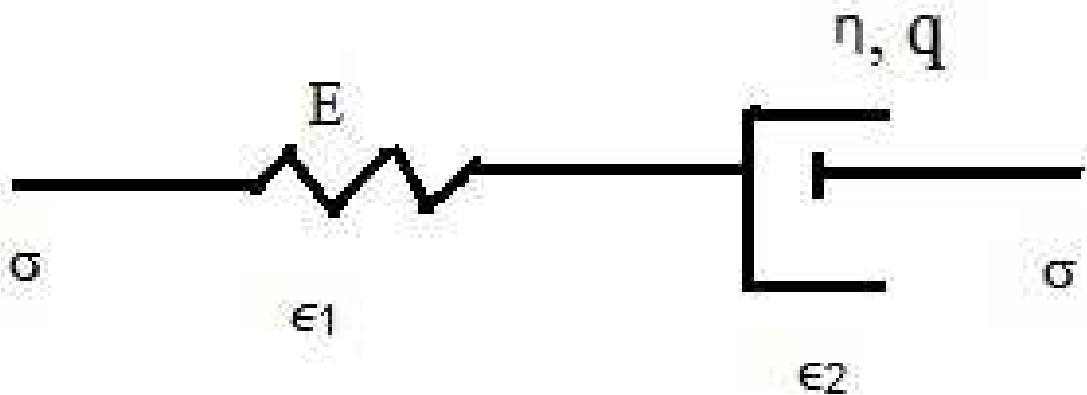}
	\end{center}
	\caption{Schematic Maxwell model. E is the spring constant and $\eta$ represents the viscosity of the fluid in the dashpot. For a stress $\sigma$, $\epsilon_1$ is the strain the the spring and $\epsilon_2$, that in the dashpot. }
	\label{maxwell}
\end{figure}

The simplest of models have at least one spring and one dashpot, respectively embodying
the elastic and viscous properties of the material. Such modelling is applicable to natural soft matter, polymer melts,
colloidal solutions etc. It may be noted, that since the models are used to study
stress-strain relations, all forces and extensions will be written as, $\sigma$ (stress) and,
$\epsilon$ (strain) respectively.  The best fit model was arrived upon after trying out various other models. Among
the simplest available models, the Maxwell model was intuitively preferred as a starting
point due to its long term fluid like nature which would better describe the more
pronounced fluid like behaviour of HPAM upon addition of salt; due to flocculation.
The Maxwell Model consists of a spring and dashpot in series as depicted in  fig.(\ref{maxwell}). 
It may be noted that the external stimulus deforms both the spring and the dashpot. The deformation of the spring is finite, but the dashpot fluid continues to deform (flow) as long as the external force acts. 

 Under a load, the stress, $\sigma$, the extension of the spring is $\epsilon_1$. Hooke's law provides:
\begin{equation}
\sigma=E \sigma_1,~~~~\label{arpr2}
\end{equation}
$\epsilon_2$ is the extension of the dashpot. From Newton's law we get,
\begin{equation}\sigma=\eta \dot{\epsilon_2}.~~~\label{arpr3}
\end{equation}
Since both elements are connected in series, the total elongation, $\epsilon$ of the system is $\epsilon=\epsilon_1 + \epsilon_2.$

Differentiating equations (\ref{arpr2}) and (\ref{arpr3}) we get the constitutive equation:
\begin{equation}
\sigma + \frac{\eta}{E} \dot{\sigma} = \eta \dot{\epsilon}~~~~\label{arpr4}
\end{equation}
It may be noted that the external stimulus deforms both the spring and the dashpot. The deformation of the spring is finite, but the dashpot fluid continues to deform (flow) as long as the external force acts. This implies that the fluid nature predominates over the solid nature and that is why this model is called the \emph{Maxwell Fluid Model.}

 Taking Laplace transform of (\ref{arpr2}) and (\ref{arpr3}) we obtain

$$\tilde{\sigma(s)}=E \tilde{\epsilon_1(s)}$$
$$\tilde{\sigma(s)}=-\eta\epsilon(0)+\eta \tilde{\epsilon_2(s)}.$$

Now at $t=0$, the material is relaxed and the strain $\epsilon(0)=0$. Therefore,
$$\tilde{\sigma}(s)=\eta s \tilde{\epsilon_2},$$
$$\tilde{\epsilon}(s)=\frac{\tilde{\sigma}}{E}+\frac{\tilde{\sigma}}{\eta s}.$$

The elastic modulus $G^*$ is defined in Laplace domain as
$$G^*=\frac{\tilde{\sigma}}{\tilde{\epsilon}}= \frac{Es\eta}{E+S\eta}.$$
In the Laplace domain, $s=i\omega,~~i=\surd(-1)$  and $\omega$ is the angular frequency of the oscillatory shear. Therefore we get
\begin{equation}
G^*(\omega)= \frac{E \eta i \omega}{E+i\eta\omega}.~~~~~~\label{arpr13}
\end{equation}
Rationalising and separating the real and imaginary parts we get
$$G^*(\omega) = G^\prime(\omega)+i G^{\prime\prime}(\omega)$$
where,
$$G^\prime(\omega)=\frac{E \eta^2 \omega^2}{E^2+\eta^2 \omega^2}$$
$$G^{\prime\prime}(\omega)=\frac{E^2 \eta \omega}{E^2+\eta^2 \omega^2}$$

Ordinarily for a perfect Hookean solid, the stress is proportional to the zeroth order time derivative of the strain and for a perfectly viscous Newtonian fluid, the stress is proportional to the strain rate i.e. the first order time derivative of the strain. We observed that linear viscoelastic models  using laws of ordinary calculus failed
to characterise the rheology of HPAM,$^{33}$ resulting in unsatisfactory theoretical fits for
the experimental curves. Therefore we formulated the constitutive relation for the dashpot, replacing the strain rate by a $q^{th}$ order time derivative of $\epsilon$, in eq.(\ref{arpr3}) where $0 < q < 1,$ signifying viscoelastic behaviour which is intermediate between that of a perfect solid and a perfect fluid as follows:
$$\sigma=\eta \frac{d^q\epsilon}{dt^q}.$$
From intuition we claim that the fractional order time derivative is characteristic of the non-Newtonian nature of complex fluids. It is to be noted that the units for viscosity remains Pa-s throughout. While introducing q into the constitutive relation for the dashpot, time is appropriately recaled by treating $\eta$ as a time dependent variable, so that the overall order of q is eliminated and the units of Pa-s. are preserved.
\\
\vspace{1cm} 

\noindent{\textbf{ Maxwell Model Reformulated}}\\
The fractional Maxwell model has been most relevant and extensively used in our work; hence the calculations of rheological quantities of interest from this model are discussed briefly.\\
\noindent{\textbf{Elastic modulus $G^*(\omega)$}}\\
\vspace{1em}
When a stress $\sigma$ is applied, strains $\epsilon_1$~and ~$\epsilon_2$ are respectively are respectively produced in the spring of spring constant E, and dashpot having coefficient of viscosity $\eta.$ From earlier discussions it follows,
\begin{equation}
\sigma(t)=E \epsilon_1 (t)~~~~~~~~~\label{arpr2010}
\end{equation}

\begin{equation}
\sigma(t)=\eta \frac{d^q \epsilon_2}{dt^q}~~~~~~~~~\label{arpr2011}
\end{equation}

Taking Laplace transform of equations (\ref{arpr2010}) and (\ref{arpr2011}), we obtain

\begin{equation}
\tilde{\sigma}(s)=E \epsilon_1 (s)~~~~~~~~~\label{arpr2012}
\end{equation}

\begin{equation}
\tilde{\sigma}(s)=\eta s^q \epsilon_2(s)~~~~~~~~~\label{arpr2013}
\end{equation}
RL type formulation requires fractional order initial states, while classical integer order initial states are needed for Caputo formulation. However, initial conditions do not give idea about history or memory. It is the integral equation or convolution operation that gives the notion of non-locality and memory. In this case, the Laplace transform of the fractional derivative is $s^q$ as the initial states are zero and thus Caputo and RL fractional derivatives are identical. The physical interpretation of the Maxwell Model and the boundary conditions that determine the solution are discussed in the \textbf{Appendix} in detail.\\

Since total strain $$\tilde{\epsilon}(s)=\tilde{\epsilon_1} (s)+\tilde{\epsilon_2} (s)$$,
 therefore,

\begin{equation}
G^*=\frac{\tilde{\sigma}(s)}{\tilde{\epsilon}(s)}=\frac{\eta E s^q}{E+\eta s^q}
\end{equation}

In the Laplace domain, $s=i\omega;$ where all symbols have their usual meaning. Now, $s^q=(i\omega)^q.$ So,
$i^q= e^{iq\pi/2}=a+ib$~ where~ $a=\cos q\pi/2$ and $b=\sin q\pi/2.$ Substituting, rearranging, rationalising the denominator and finally separating real and imaginary parts we obtain,

\begin{equation}
G^*(\omega)=G^{\prime}(\omega)+i G^{\prime\prime}(\omega)
\end{equation}

where
\begin{equation}
G^{\prime}(\omega)=\frac{Ea+\eta \omega^q}{\frac{E}{\eta\omega^q}+2a+\frac{\eta\omega^q}{E}}
\end{equation}

\begin{equation}
G^{\prime\prime}(\omega)=\frac{Eb}{\frac{E}{\eta\omega^q}+2a+\frac{\eta\omega^q}{E}}
\end{equation}

are respectively the storage and loss modulus.
The frequency dependence of $G^{\prime}$  and $G^{\prime\prime}$ gives a description of the rheology of a material.$^{33}$ For
many bio-polymers, $G^{\prime\prime}$ exceeds $G^{\prime}$ at low frequencies. This indicates a fluid like
response where the material flows, but for higher frequencies, $G^{\prime}$ dominates $G^{\prime\prime}$,
implying that the molecules of the material cannot follow the rapid oscillations in
strain and the sample responds like a solid. The loss modulus usually exhibits a
peak at a certain frequency, the inverse of this loss-peak frequency represents the
relaxation time of the sample.

Another rheological property of interest is the \textbf{complex viscosity} defined by
\begin{equation}
\eta^*=\eta^\prime - i \eta^{\prime\prime}=\frac{G^*(\omega)}{i~\omega}.~~~\label{arpr52}
\end{equation}
Its magnitude is given by
\begin{equation}
\eta^*(\omega)= \frac{\surd( {G^\prime}^2 + {G^{\prime\prime}}^2)}{\omega};~~~\label{arpr53}
\end{equation}
so that $\eta^\prime=G\prime/\omega$ and $\eta^{\prime\prime}=G^{\prime\prime}/\omega$. The complex viscosity $\eta^*(\omega)$ can be calculated from eq.(\ref{arpr53}).
\vspace{1cm}

\noindent{\textbf{Creep compliance - J(t)}}
\vspace{1em}

The creep compliance is another quantity of rheological interest pertaining to the creep phase of the standard test. It is called compliance, as it is
the inverse of modulus and is a transfer function relating stress and strain as:
$$J(t)=\frac{\epsilon(t)}{\sigma_0}$$
where all notations have usual meanings.

In the Laplace domain, $\tilde{J}(s)$ is defined as
$$\tilde{J}(s)=\frac{\tilde{\epsilon}(s)}{\tilde{\sigma}(s)}=\frac{1}{E}+\frac{1}{\eta s^q}.$$
We perform inverse Laplace transform over $\tilde{J}(s)$ to obtain creep compliance $J(t)$ in the time domain. Thus
$$J(t)= \mathcal L^{-1}\{\tilde{J}(s)\}= \frac{1}{E}\mathcal L^{-1}(1)+\frac{1}{\eta}\mathcal L^{-1}(\frac{1}{s^q})$$
This gives 

$$J(t)=\frac{1}{E}\delta(t)+\frac{1}{\eta}\frac{t^{q-1}}{\Gamma(q)}$$
where $\delta(t)$ is Dirac delta function and $\Gamma(q)$ denotes the Gamma Function.$^{34}$

 \section*{Results and Analysis}
 \subsection*{Effect of salinity}
Zhang et. al. conducted the rheological experiments investigating  oscillatory and flow shear behaviours of hydrolysed polyacrylamide (HPAM).$^{30}$ They observed the variation of  storage  modulus $G^\prime (\omega)$ in units of pascal (Pa) as well as loss modulus $G^{\prime\prime}(\omega)$ in units of pascal (Pa) with angular frequency $\omega$ in units of (rad/s) through oscillatory shear tests at five different salinities (0 mg/L, 250 mg/L, 500 mg/L, 3400 mg/L, 6800 mg/L, keeping the solution concentration of HPAM fixed at 1000 mg/L.

Zhang et. al. also performed flow shear tests investigating the variation of shear viscosity $\eta_{app}$ in units of pascal-second (Pa-s) with shear rate,$^{30}$ $\dot{\gamma}$ in units of $s^{-1}$.

 The length of the chain of polymer molecules is the main factor determining its rheology. Polymeric liquids having microstructure can develop anisotropy in the orientation of the constituent polymers in flow, thereby leading to normal stress differences which are zero for an isotropic liquid. Extensional flow is where the local kinematics dictates that the fluid element is stretched in one or more directions and compressed in others. The empirical Cox-Merz rule states that the apparent viscosity,~ $\eta_{app}$~ should be the same function of shear rate $\dot{\gamma}$ as $|\eta^*|$ is of frequency $\omega$,$^{35}$ where
 \begin{equation}
 \eta^*= \frac{\surd({G^\prime}^2 + {G^{\prime\prime}})}{\omega}^2
 \label{cox}
 \end{equation}
   This rule is valid for polymer melts as the flow shear curves and oscillatory shear curves are almost identical, below exceptionally high frequenices. This rule was applied while modelling the flow shear data in terms of complex viscosity.
Applying the Cox-Merz rule we have modelled the complex viscosity $\eta^{*} (\omega)$ which is a function of angular frequency $\omega,$ in (rad/s) from eq.(\ref{cox}), using the data obtained from shear flow tests. The domain of validity of this rule falls in our area of application.$^{35}$ 

\begin{figure}[H]
	\begin{center}
	\includegraphics[width=15.2cm]{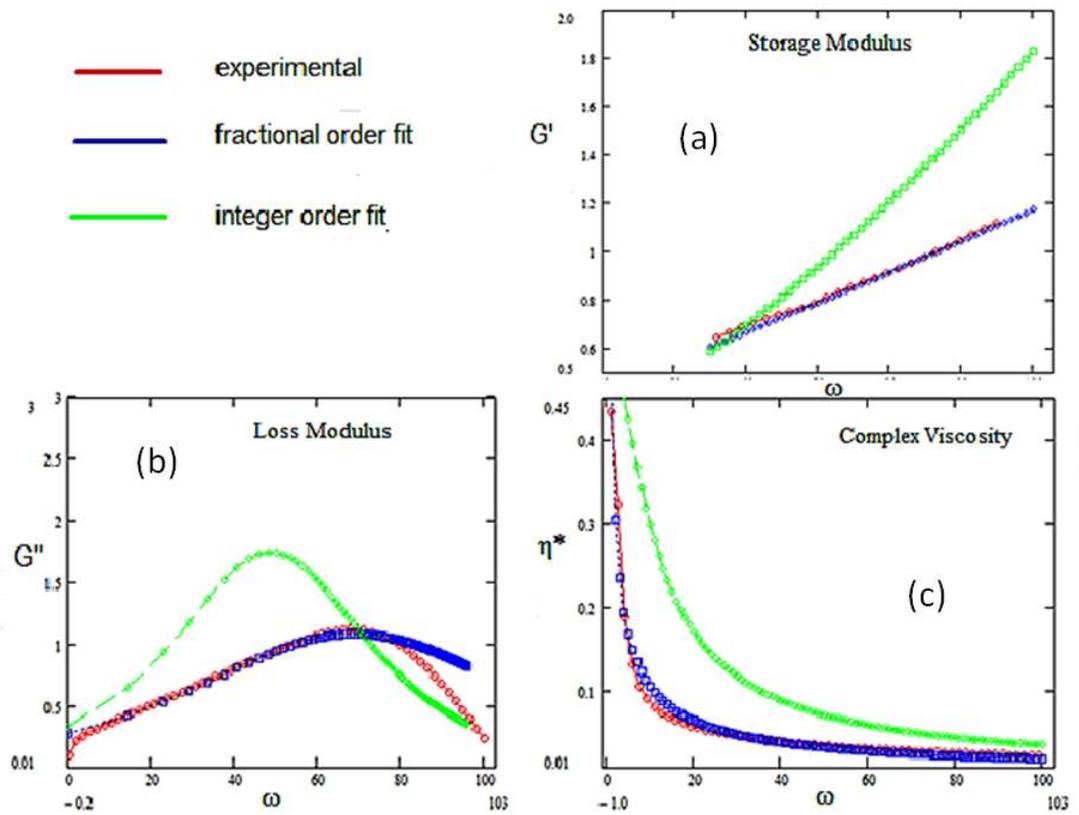}
	\end{center}
\caption{The rheological properties of HPAM obtained from experimental data compared with calculated values using Maxwell's Model using integer derivative and fractional derivative calculus.$^{30}$ (a) Storage modulus (b)Loss modulus (c)Complex viscosity. In all cases, the fractional derivative calculus fitted best the experimental data. }
\label{integfrac}	
\end{figure}

We observed that linear viscoelastic models using laws of ordinary calculus failed to characterise the rheology of HPAM, resulting in unsatisfactory theoretical fits for the experimental curves, as demonstrated by the graphs, fig.(\ref{integfrac}).

\begin{figure}[H]
	\centering
	\includegraphics[width=15.2cm]{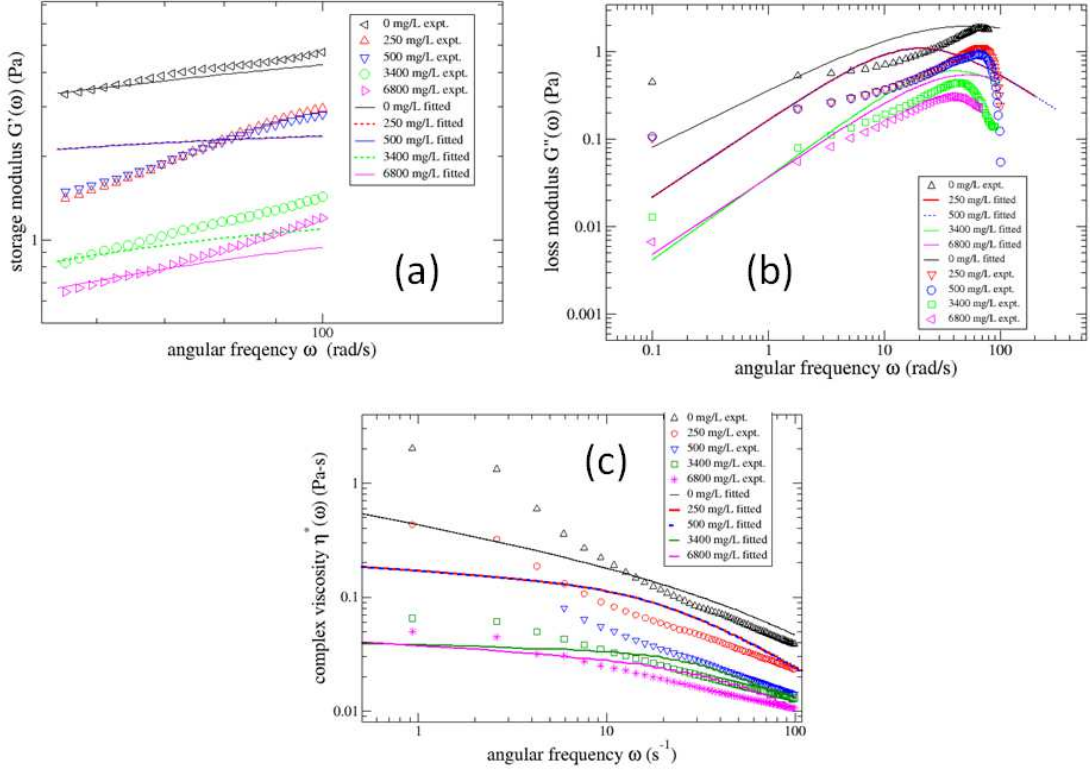}
	\caption{Salinity variation:(a)Comparison between experimental and calculated storage modulus using Maxwell Model.$^{30}$ (b)Comparison between experimental and calculated loss modulus using Maxwell Model.$^{30}$ (c)Comparison between experimental and calculated complex viscosity using Maxwell Model.$^{30}$}
	\label{Bestfitpanel}
\end{figure}
But when the same was fitted using expressions calculated using a fractional order derivative  the modelling was successful resulting in agreeable theoretical extrapolation of the experimental curves. This led us to verify what we expected - the fractional order q, introduced into the constitutive relation for the dashpot embodies the non-Newtonian nature of HPAM. The dashpot in itself provides an insufficient description of a complex fluid like HPAM as it only embodies a simple fluid obeying Newton's law of viscosity. This validates the reformulation of constitutive equations using the techniques of fractional calculus, thereby highlighting its indispensability in rheological studies.
The two element fractional Maxwell Model with parameters $E,~ \eta,$ and q was concluded to be the best fit model.  Figure.(\ref{Bestfitpanel}) demonstrate the best theoretical fits of rheological properties of HPAM for experimental curves obtained for the five different salinities. The best fit values of the parameters $E,~ \eta,$ and q for all salinities are displayed in Table.(\ref{salttable}).
\begin{table}[h]
\begin{center}
\caption{Model parameters for salinity variation of $\eta^{*}$}
\vspace{0.5cm}
\begin{tabular}{|c|c|c|c|}
\hline  Salinity(mg/L)& E & $\eta$ & q \\
\hline 0 & 6.72 & 0.445 & 0.678 \\
\hline 250 & 2.58 & 0.172 & 0.896 \\
\hline 500 & 2.58 & 0.172 & 0.815 \\
\hline
\hline 3400 & 1.298 & 0.038 & 0.96 \\
\hline 6800 & 1.298 & 0.038 & 0.89 \\
\hline
\end{tabular}
\label{salttable}
\end{center}
\end{table}

 The software used for modelling was Mathcad Professional 2001. Origin Pro 8 was used to extract data. Wxmaxima was used to predict the curves for creep compliance.

Though the Maxwell model was concluded to be the best fit model governing the behaviour of the sample at all salinities, it was seen that one set of fixed $E, \eta$ values characterised the salinities 250mg/L and 500mg/L while varying q. Similarly, another set of constant $E,~ \eta$ values characterised the salinities 3400mg/L and 6800mg/L, while varying q. The two sets of $E, \eta$ values characterising two groups of salinities indicate two classes of viscoelasticity displayed by the given sample. It is evident that when the order of the salinity changes by a factor of 10, there is a drastic change of  the  viscoelastic properties of the sample although the governing model remains the same. The addition of $Na^{+}$ effectively screens
the negative charges  on carboxyl groups. This reduces the electrostatic
repulsion within the polymer chains. Thus conformational
transition of the polymer from a stretched state to a shrinkable
state decreases the hydraulic radius of the chain and the degree
of polymer chain entanglement, resulting in reduction of viscosity
of polymer solution.$^{36-38}$ Thus the complex viscosity $\eta^{*}$ decreases rapidly with increasing
salt concentration. Jung et al. have also shown experimentally that the addition of 
NaCl into HPAM solution significantly
reduced the polymer solution viscosity.$^{39}$ Lopes et al. have argued that it is not only the amount of salt present in the core,$^{40}$ but also the type of cation present in the salt, that influences HPAM properties. Keeping the amount of salt added and polymer concentration fixed, they have observed experimentally, that a salt having a trivalent cation such as $Al^{3+}$  reacts most readily with anionic part of the polymer chain, causing maximum flocculation; followed by divalent cations like $Ca^{2+}$, and finally $Na^{+}$ having a comparatively weaker flocculating potential. This flocculation leads to reduction in the apparent viscosity of HPAM.
\begin{figure}[H]
\begin{center}
\includegraphics[width=15.2cm]{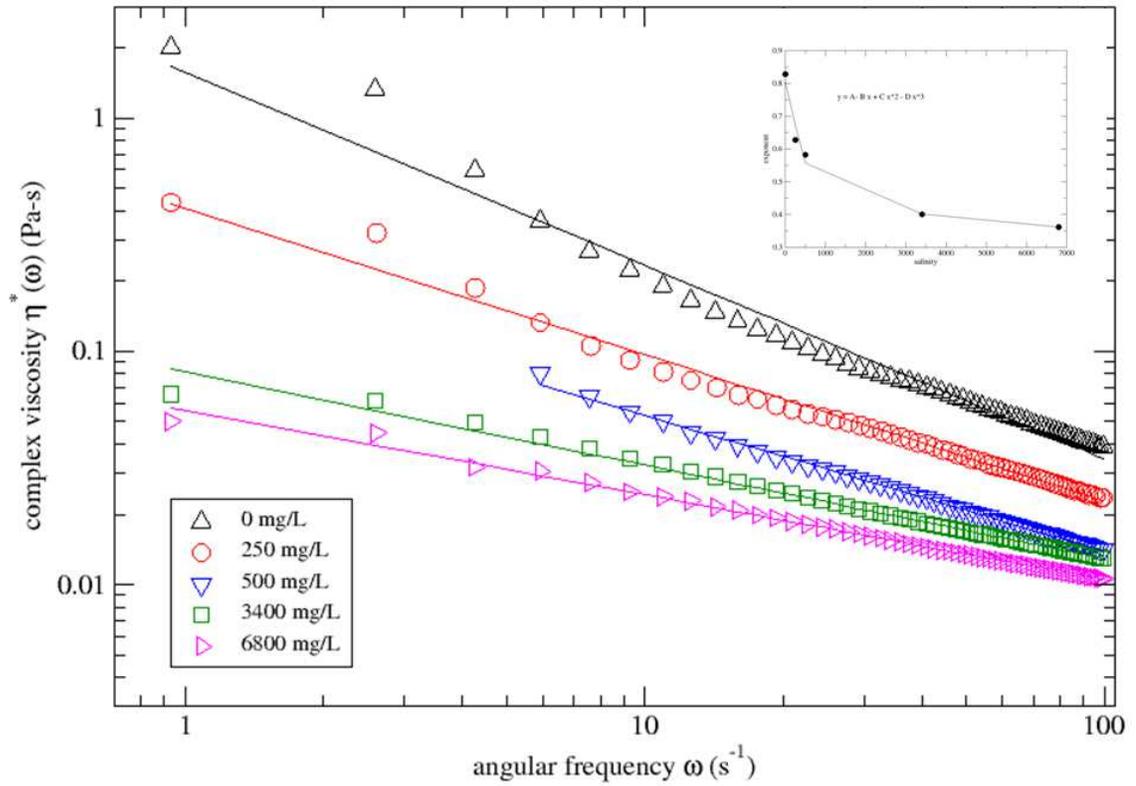}
\end{center}
\caption{Log-log plot of complex viscosity versus angular velocity. The inset shows the cubic dependence of the exponent on salinity}
\label{powerlaw}
\end{figure}

An important observation is that although best fits are obtained for values of q close to 1; at q=1 the model fails to characterise the rheology of HPAM, implying that the polymer melt is indeed non-Newtonian. The fitted values differ significantly from 0 mg/L to 250 mg/L indicating that addition of salt severely changes the viscoelasticity of the sample.

When the complex viscosity obtained from our calculations when plotted against the angular velocity obtained from Zhang et al,$^{30}$ a power law fit was obtained as shown in fig.(\ref{powerlaw}). The exponent for different salinities followed a cubic fit as shown in the inset of fig.(\ref{powerlaw}.)

\subsection*{Prediction of creep compliance}
From the best fit Maxwell model, and using  the fitted parameter values, other rheological quantities of interest may be predicted. One such prediction is that of the creep compliance given by
$$J(t)=\frac{\epsilon(t)}{\sigma_0};$$
where $\sigma_0$ is the applied constant stress. 
\begin{figure}[H]
\begin{center}
\includegraphics[width=15.2cm]{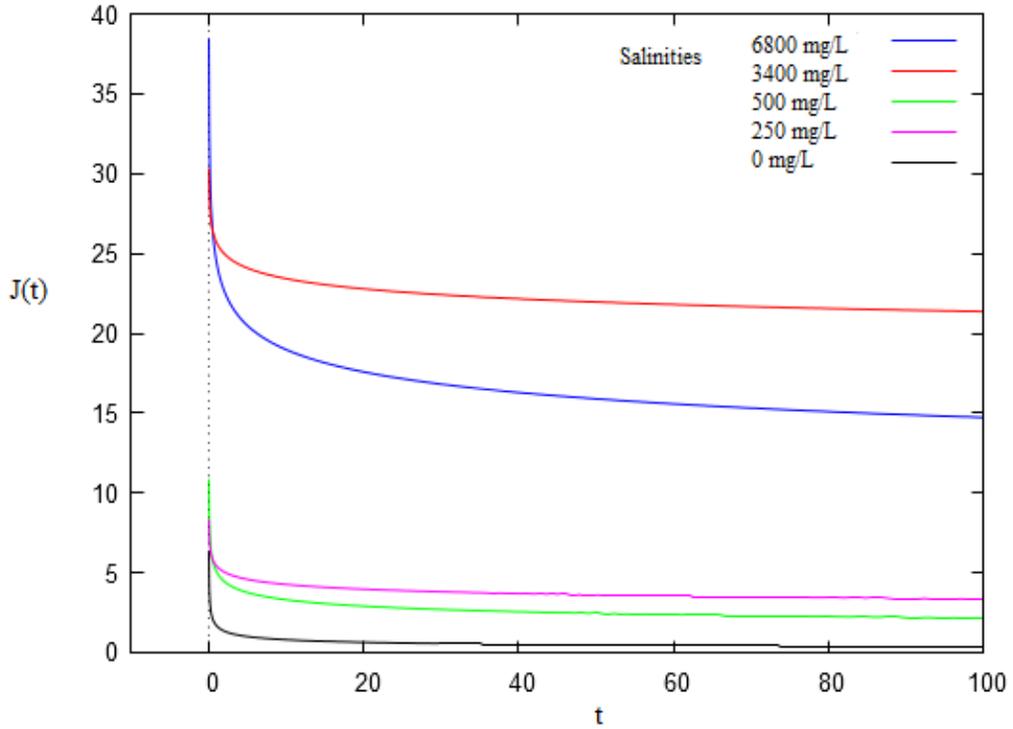}
\end{center}
\caption{Variation of creep compliance with time for different salinities as predicted from theory }
\label{Creep}
\end{figure}
\begin{table}[h]
\begin{center}
\caption{Model parameters for calculation of creep compliance}
\vspace{0.5cm}
\begin{tabular}{|c|c|c|c|}
\hline Salinity (mg/L)& q (fractional order) & $\Gamma(q)$&1-q  \\
\hline 6800& 0.890 & 1.07683 & 0.11\\
\hline 3400& 0.960 & 1.02473 & 0.04\\
\hline 500 & 0.815 & 1.14779 & 0.185 \\
\hline 250 & 0.896 & 1.07188 & 0.104 \\
\hline 0 & 0.678 & 1.3343 & 0.322\\
\hline
\end{tabular}
\label{creeptable}
\end{center}
\end{table}

The general expression for creep calculated from the Maxwell model is given by
\begin{equation}
J(t) = \frac{1}{E} \delta(t) + 1/[\eta \Gamma(q)t^{(1-q)}].
\label{creepeq}	
\end{equation}

Putting values of $E, \eta, q $ and  $\Gamma(q)$ from Tables.(\ref{salttable}) and (\ref{creeptable}) into eq.(\ref{creepeq}), the creep compliance has been predicted for different salinities as shown in fig.(\ref{Creep}).

The nature of the curves predicted for creep compliance is in agreement with the conclusions that were drawn from the creep phase of the \textit{Standard Test} performed on the Maxwell model. A sudden loading at $t=0$ is well characterised by Dirac delta function in the calculated expression. The curves blow up at t=0. There is a gradual decrease of slope with time. Also it is evident from the curves that there is a clear bunching of salinity dependent properties of the sample for each viscoelastic class.
The slope of the higher of the two salinities of a particular viscoelastic class decreases more rapidly than the corresponding lower salinity. Physically this is explained as follows: we know that ability to deform infinitely under finite stress ie, the ability to creep is more for a fluid and is almost absent for a solid. The HPAM structure may be visualised as a linear crystalline chain of molecules (not perfectly rigid owing to its viscoelastic nature). Since it is an anionic polyelectrolyte, the negative ions at ends repel each other, upholding their geometry. On adding salt, the salt cation reacts with the anionic ends leading to flocculation, destroying the linear crystal like geometry. In this state, in the solution, the fluid like character becomes more pronounced with increasing salinity. So its ability to creep increases. Therefore, at a given stress, $\sigma_0$, the creep compliance decreases more rapidly for a higher salinity as the corresponding strain is dissipated more rapidly.

\subsection*{Effect of Temperature}\label{sectemp}
The testing temperature at which the experiments were performed by Zhang et al. was $25^{o}$C.$^{30}$
However, along with salinity, a high reservoir temperature $\sim$ $70^{o}$C also affects
HPAM properties. 

\begin{figure}[H]
\begin{center}
\includegraphics[width=15.2cm]{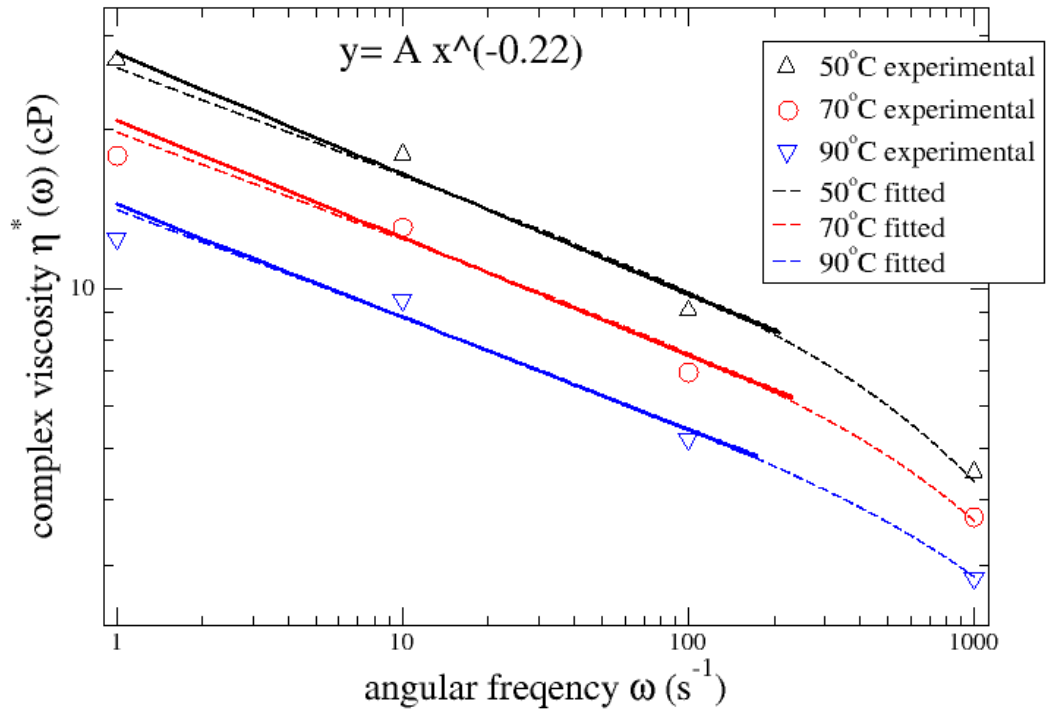}
\end{center}	
\caption{Variation of complex viscosity versus angular frequency for different temperatures: comparison between experiment and theory.$^{31}$ For angular frequencies $\sim$120 $s^{-1}$, the variation follows a power law with exponent of 0.22 for every temperature. The power law fit is shown by thick solid line.}	
\label{temp}
\end{figure}

Gao et al. measured the complex viscosity of HPAM at different temperatures:$^{31}$ $50^{o}$C, $70^{o}$C and at $90^{o}$C, keeping a constant concentration, pH and at zero salinity. We used the Maxwell model described above to calculate the complex viscosity. The model parameters are displayed in Table.(\ref{temptable}), where it is evident that the spring constant $E$ and the order of the fractional derivative $q$ were identical for each temperature. The parameter $\eta$ was tuned to match experimental values. Fig.(\ref{temp}) shows the comparison between experimental and calculated values of $\eta^{*}$ at different angular velocities, the match between experimental and theoretical values for each temperature being very good. At any particular angular frequency, $\eta^{*}$ decreases with increase in temperature as the long chain molecules of HPAM tend to disentangle. The reason for the decrease in $\eta^{*}$  is attributed to the intensification of the thermal motion of the hydrophobic group, in the polymer chain. Thus the hydration
of the hydrophobic group is weakened, which in turn reduces the hydrodynamic volume with increasing temperature.

From the model parameters in Table(\ref{temptable}), it is observed that the viscous coefficient of the dashpot fluid increases with decreasing temperature. This is indicative of a more pronounced viscous behaviour of polymer chains at lower temperatures and substantiates our previous arguments regarding the same.
Another point of interest is the power law exponent at $\alpha=0.22$ for each of the fits, regardless of temperature, upto a shear of 120 $s^-1$ only. This value of shear is thus indicative of a point of characteristic transition between Newtonian and non-Newtonian behaviour, which is indicated by a sudden drop for each curve at this particular of shear as shown in fig.(\ref{temp}).

\begin{table}[h]
\begin{center}
\caption{Model parameters for temperature variation of $\eta^{*}$}
\vspace{0.5cm}
\begin{tabular}{|c|c|c|c|}
\hline  Temp.($^o$C ) & E & $\eta$ & q \\
\hline  90 & 7588 & 14.08 & 0.798\\
\hline  70 & 7588 & 19.81 & 0.798\\
\hline  50 & 7588 & 26.18 & 0.798 \\
\hline 
\end{tabular}
\label{temptable}
\end{center}
\end{table}

\subsection*{Role of Concentration and pH} 
\begin{figure}[H]
\begin{center}
\includegraphics[width=15.2cm]{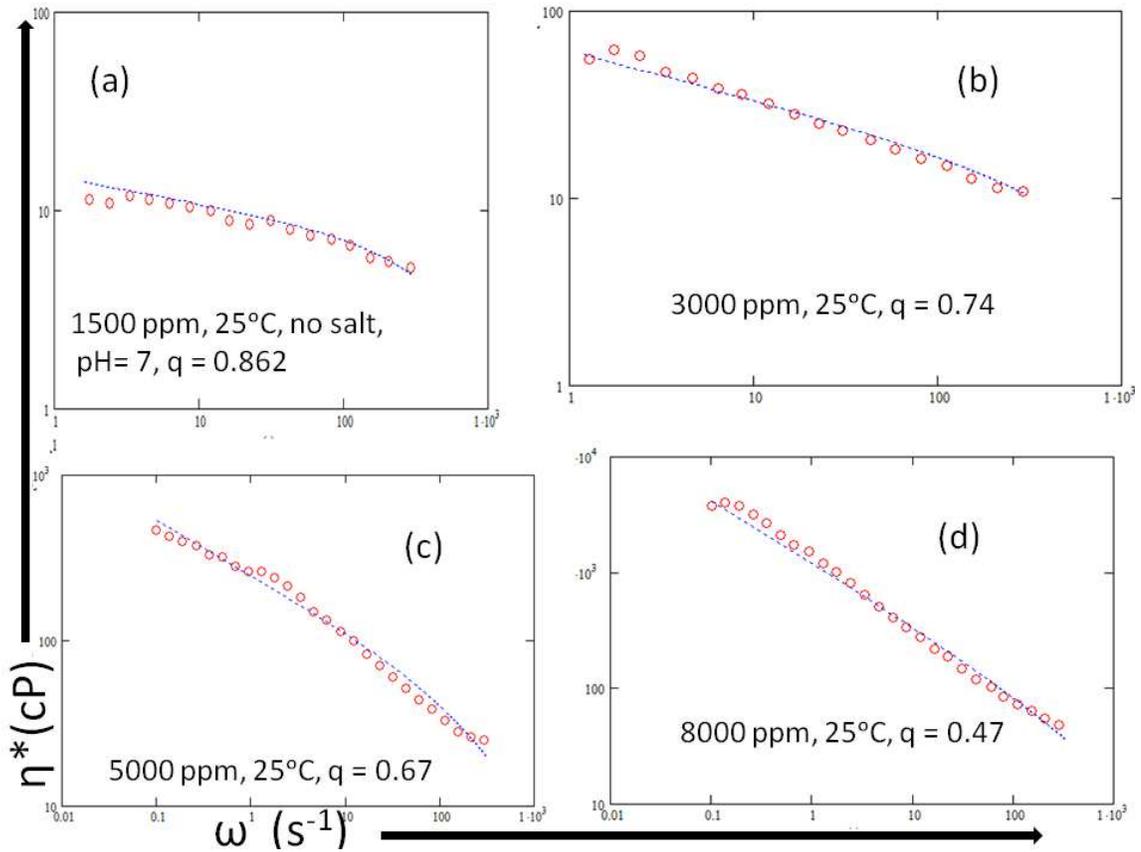}
\end{center}
\caption{Variation of complex viscosity versus angular frequency for different polymer concentrations at constant temperature of 25$^{o}$C, pH=7 and zero salinity, (a)1500 ppm (b)3000 ppm (c) 5000 ppm (d) 8000 ppm. Comparison between experiment and theory is good in each case.$^{32}$ The order of the fractional derivative decreases with increase in concentration as displayed in the graph.}	
\label{conc}
\end{figure} 
\begin{figure}[H]
\begin{center}
\includegraphics[width=15.2cm]{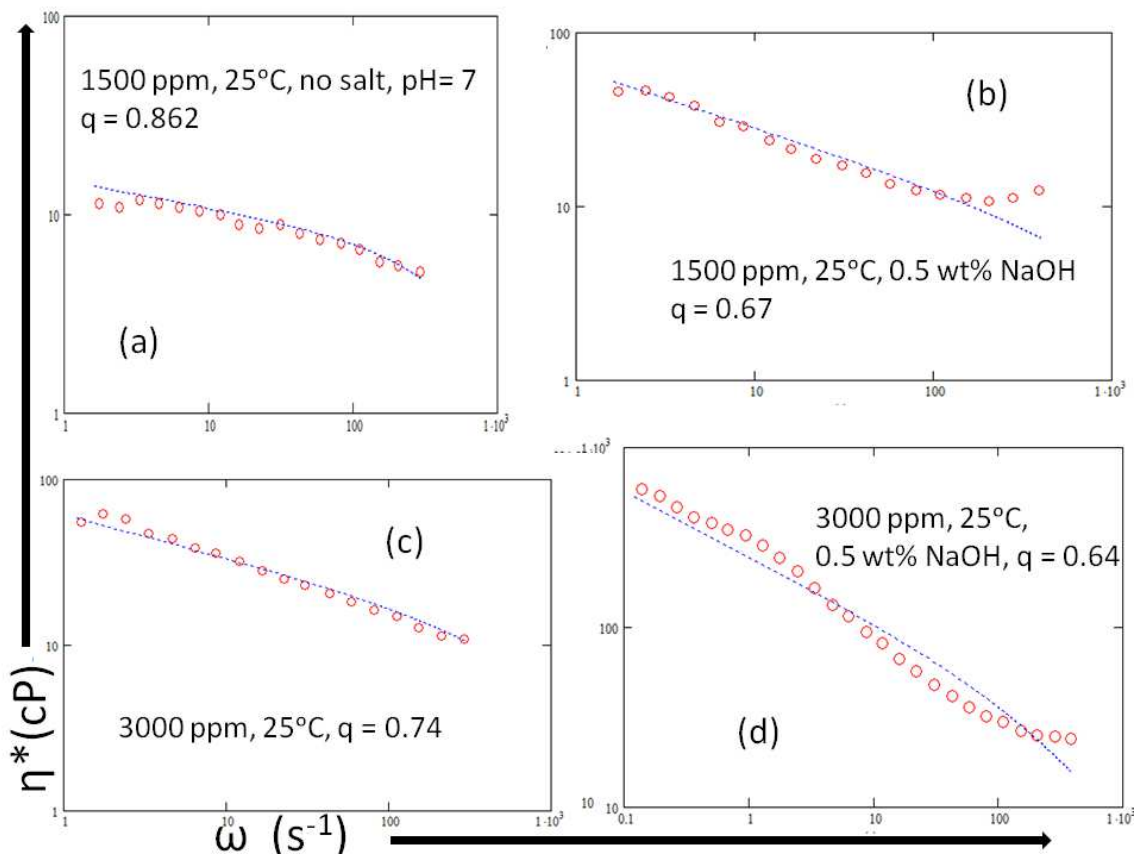}
\end{center}	
\caption{Variation of complex viscosity versus angular frequency for different pH. Experimental data from have been modelled.$^{32}$ (a) and (b) show the effect of pH for a polymer concentration of 1500 ppm at constant temperature of 25$^{o}$C and zero salinity. (c)and (d)show the effect of pH for a polymer concentration of 3000 ppm at constant temperature of 25$^{o}$C and zero salinity.  The order of the fractional derivative decreases with increase in the alkalinity.}	
\label{pH}
\end{figure}
At very low concentrations, there is no strong interaction between polymer molecules. Physical properties change in direct proportion to the concentration. With increasing concentration the increment in viscosity increases at a faster rate. At concentrations below the critical concentration $C_p$, polymer molecules in the solution have intra-molecular association, and the molecular chains tend to shrink, resulting in a smaller $\eta^{*}$. When the concentration reaches $C_p$, $\eta^{*}$ increases sharply because the solution has a supra-molecular agglomerate
structure that enlarges the hydrodynamic volume at and above $C_p$.
This results in a notable increase in $\eta^{*}$, fig.(\ref{conc}).
It may be noted, that the rheology of polymer melts is polymer specific and is characterised by the particular concentration, molecular weight and temperature. 

The shear viscosity of HPAM is strongly dependent on the variation
of pH due to the pH-sensitive carboxyl groups along its
backbone chain. Jung et al. have demonstrated experimentally that the viscosity of HPAM increases with NaOH concentration
at lower alkaline pH.$^{39}$  With increase in alkalinity, the viscosity slightly declines and
ultimately becomes steady at higher concentration.  
On increasing the alkalinity, the negatively
charged carboxyl groups along the chains make the polymer
molecules tend to coil up due to the electric repulsion. This increases the hydrodynamic radius which leads to a large increase in solution viscosity. Our theoretical calculation of $\eta^{*}$ using the Maxwell model and for two different polymer concentrations, follow the experimental findings of Jung et al. quite well, fig.(\ref{pH}).

The complex viscosity increases sharply with increase in alkalinity, for both the polymer concentrations. It is to be noted that the order of the fractional derivative q, decreases with increase in pH in both cases. 
\section*{Discussion}
It is clear that the effective rheological properties of polymers like HPAM are affected by concentration, presence of salt, temperature and pH. The question naturally arises whether it is possible to tailor a rheological property of interest , eg. complex viscosity, for a certain shear rate to suit a particular purpose. Any combination of the parameters that offset each other's effect on the rheological property of the polymer solution may be termed as \textit{ complementary conjugates.}
\subsection*{Tailoring rheology of HPAM}
\begin{figure}[H]
\begin{center}
\includegraphics[width=15.2cm]{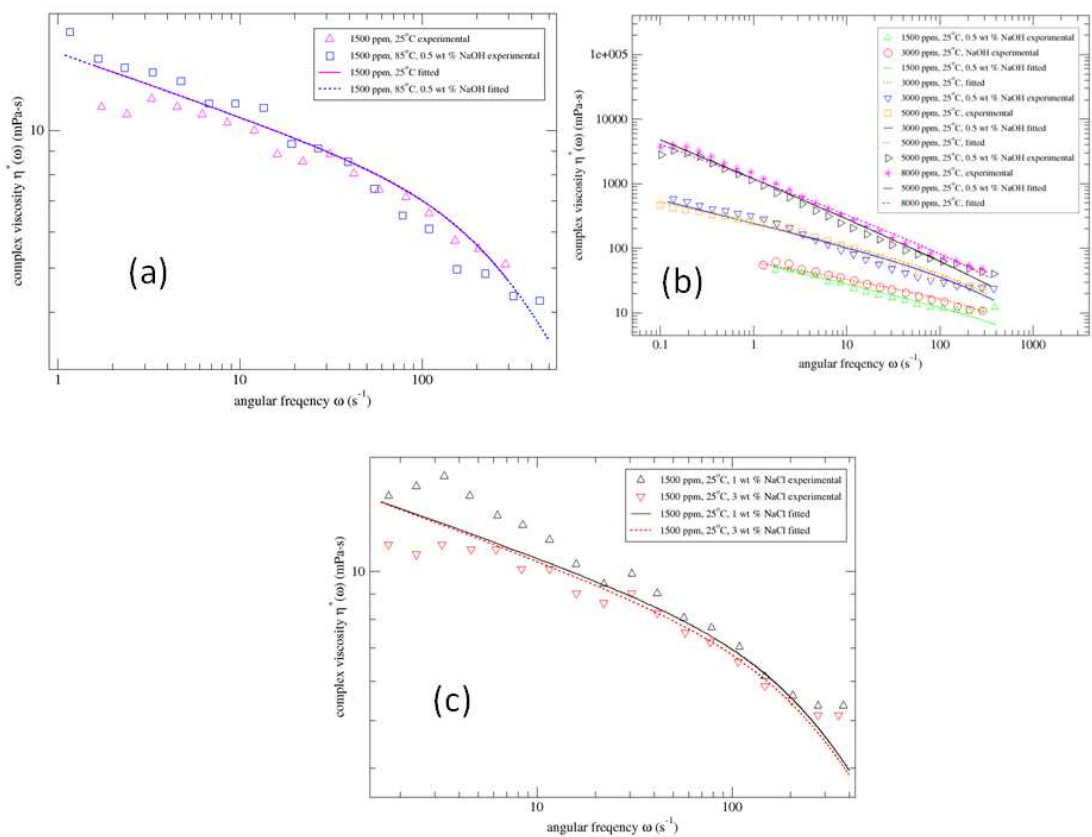}
\end{center}	
\caption{Variation of complex viscosity versus angular frequency for different combinations of \textit{complementary conjugates}. Experimental data from Choi e.t al. have been modelled.$^{32}$ (a)Higher temperature is offset by a suitable increase in pH. (b)Effect of increased polymer concentration offset by lower the pH value. (c)Variations in salt concentration alone cannot offset small changes in concentration.}	
\label{combination}
\end{figure}
A careful examination of fig.(\ref{combination} a) reveals that at any given polymer concentration, here 1500 ppm, the effect of a higher temperature on effective complex viscosity, is offset by a suitable increase in pH. This also follows from the results of figs.(\ref{temp}) and (\ref{pH}). In this manner it is possible to match the complex viscosity at a similar shear rate to the case of a neutral solution of HPAM at a lower temperature. Thus temperature and pH are a set of `complementary conjugates'.

In a similar manner, it is possible to label polymer concentration and pH of the solution as a second set of `complementary conjugates'. Figure.(\ref{combination} b) shows the effectiveness of offsetting the effect of an increased polymer concentration by lowering the pH value; in the cases studied, the higher concentrated polymer solution was maintained neutral. We have demonstrated this for the different combinations of polymer concentrations.

In fig.(\ref{combination} c), the only variable is the salt concentration which does not greatly affect the viscosity for small changes in concentration.
\\
\begin{table}[h]
\begin{center}
\caption{Model parameters for tailoring $\eta^{*}$ using polymer concentration, pH and salinity.}
\vspace{0.5cm}
\begin{tabular}{|c|c|c|c|c|c|c|c|c|}
\hline Type&C(ppm)&T($^o$C)&NaCl(wt $\%$)&NaOH(wt$\%$)&$\eta^{*}$(mPa)& $\omega$(s$^{-1}$) &q& $\alpha$\\
\hline  I   & 1500 & 25 & 0 & 0   & 34.94 & 0.002 & 0.86 & 0.17\\
\hline  I   & 1500 & 85 & 0 & 0.5 & 34.94 & 0.002 & 0.86 & 0.17\\
\hline  II  & 1500 & 25 & 0 & 0.5 & 34.10 & 5.6   & 0.67 & 0.25\\
\hline  II  & 3000 & 25 & 0 & 0   & 34.74 & 8.2   & 0.74 & 0.23 \\
\hline  II  & 3000 & 25 & 0 & 0.5 & 34.74 & 105.1 & 0.64 & 0.45\\
\hline  II  & 5000 & 25 & 0 & 0   & 34.70 & 128.1 & 0.67 & 0.42\\
\hline  II  & 5000 & 25 & 0 & 0.5 & 34.66 & 235.1 & 0.40 & 0.60\\
\hline  II  & 8000 & 25 & 0 & 0   & 34.63 & 349.1 & 0.47 & 0.62\\
\hline  III & 1500 & 25 & 1 & 0   & 34.69 & 0.002 & 0.86 & 0.17\\
\hline  III & 1500 & 25 & 3 & 0   & 34.61 & 0.003 & 0.84 & 0.17\\
\hline
\end{tabular}
\label{table4}
\end{center}
\end{table}
\\

Table.(\ref{table4}) displays the values of the model parameters used to match experimental data.  Types (I) and (II) highlight the efficiency of the \textit{complementary conjugates},temperature and pH, and, concentration and pH,  in tailoring the rheological behaviour of polymer melts to a desired value, eg. $\sim$34 mPa.  Type(I) case achieves this value at an angular frequency of $0.002$s$^{-1}$, while Type(II) variants do so at different other values  of $\omega$ as shown in the Table. Type (III) , which shows the effect of only one parameter (salinity), does not affect $\eta^{*}$ much.

What is the significance of $q$ as a parameter in our model? The answer becomes apparent when one examines the last three columns of Table.(\ref{table4}). For each pair of `complementary conjugates', the order of the fractional derivative of the Maxwell's model are $\sim$ equal, as is the corresponding angular frequency $\omega$. Further as $\omega$ increases, $q$ decreases.
It may be expected that with low $\omega$, if a particular $\eta^{*}$ is to be achieved, the visco-thickening will take longer time to develop leading to a low power law exponent $\alpha$. The 'memory' of the entire process is expected to be long with a larger value of the order $q$. The reverse argument is true for a larger shear velocity $\omega$, leading to a sharper change in  $\eta^{*}$ reflected in higher $\alpha$ exponent of the power law. This in turn results in a shorter memory ($q$) of the dynamical development of $\eta^{*}$.

 \section*{Conclusions} \label{conclusions}
 We have demonstrated that the two-element Maxwell model is appropriate to study the rheology of polymer melts like HPAM, using fractional calculus in the constitutive equations. The model was successful in reproducing experimental data on complex viscosity reported by several other groups for different parametric variations. The role of variations in salinity, temperature, concentration and pH on complex viscosity of the polymer melt HPAM was extensively studied. Complex viscosity of HPAM changes dramatically by several orders of magnitude for relatively small changes in pH, polymer concentrations and salinity. This is attributed to the suppression of repulsion between the carboxyl groups in the polymeric chains which can be controlled by pH and addition of appropriate salts. 
This study concludes that HPAM shows a long term fluid like behaviour exhibiting two different classes of viscoelasticity at low and high salinity ranges respectively. The modelling also predicts an intimate correlation of the salinity dependence of rheological properties, viz, creep compliance with each viscoelastic class. One of the most important uses of HPAM is in the oil recovery process.$^{41}$ For polymer flooding purposes, as far as achieving the target viscosity of HPAM is concerned, the salinity of the reservoir becomes the dominating factor.

 The role of temperature on complex viscosity was investigated and matched the reported experimental studies. $\eta^{*}$ was found to decrease with increase in temperature. This behaviour is attributed to the intensification of the thermal motion of the hydrophobic group, in the polymer chain, which weakens the hydrophobic group. This in turn reduces the hydrodynamic volume with increasing temperature. At all temperatures studies, HPAM showed a power-law behaviour up-to a critical value of 120 $s^-1$, after which the viscosity rapidly fell with further increase in shear. We propose that this critical value of shear is indicative of a characteristic change in the viscosity, perhaps from viscoelastic property to non-Newtonian behaviour.
 
 The model was successful in predicting the creep compliance of HPAM as the behaviour of the polymer melt matched our intuitive expectation of such viscoelastic fluids. 
This success is much more than mere curve fitting to experimental data as the derivation comes from Causality principle where the kernel function is a decaying power law. This gives an idea of memory relaxation. 
 We have demonstrated that complex viscoelastic fluids show `memory' effects that determine their emerging dynamics. Different perturbations to the polymer melts, affect their `memory' differently leading to different power-law behaviour between the same pair of variables. The variation of $\eta^{*}$ with angular frequency $\omega$ showed a power law behaviour for almost all variations in parameters. Power law dependence between two variables is observed in `emerging' phenomena that show a \textit{wide distribution}. It is not surprising therefore that complex viscosity in viscoelastic materials will show this nature as they display creep long after stress is removed. Thus the lower the exponent of the power law behaviour, the longer the `memory' of its past is retained, as viscoelastic fluids show large Deborrah Number $De$, i.e. a large relaxation time. Our modelling using fractional calculus essentially made this connection as the fractional order derivative $q$ which is now recognised as a memory index to emergent phenomena, showed an inverse relationship with respect to the power-law exponent $\alpha$; - the higher the memory index $q$, the smaller is the power-law exponent.

 Lastly, we have demonstrated that it is possible to tailor the complex viscosity of HPAM by suitably tuning `complementary conjugates', in order to meet requirements of industry and technology. We have predicted two pairs of `complementary conjugates' for HPAM: (i)temperature and pH (ii)concentration and pH.
 
 One of the most important uses of HPAM is in the oil recovery process being easily available and highly cost effective.$^{41}$  It is resistant to bio-degradation, has good viscosifying properties and reduces water permeability. The high viscosity of HPAM is responsible for improving the mobility ratio and water injection profile which enhances the oil recovery ratio. HPAM is also widely used in food products, paper
manufacturing, mining, waste-water treatment, and in the isolation of enzymes. The judicious tuning of the `complementary conjugates' may result in HPAM with suitable rheological properties that will enhance its effectiveness in each of these applications.

 Before we conclude, it seems essential to include a brief discussion on what we mean by `memory index to emergent phenomena' and other perspectives it opens up on the same problem, for completeness. In condensed matter theory, the notion of associating emergent phenomena to the spontaneous breaking of a continuous symmetry is an exciting area of research. From intuition, we presume the existence of a symmetry in the polymer chains constituting HPAM in the Newtonian limit which breaks in the non linear regime leading to memory retention displayed by viscoelastic materials, which is the emergent phenomena. Through our constitutive relations, we use fractional derivatives of order q to characterise the dependence of the system on its stress history, q being directly proportional to the memory retention capacity of a system. For example, theoretically $q = 0$ for Markov processes.  In generalised calculus, the process of differintegration is equivalent to a two stage memory process; the Standard Test being the viscoelastic analogue of such a mechanism.$^{14}$ Whenever the viscoelastic material is subjected to an external perturbation through one of the stressing parameters (viz. temperature, pH etc.) the power law dependence is a manifestation of the emergent phenomena. In section(\ref{sectemp}) for example, the independence of the power law exponent of temperature in the non-Newtonian regime, is a quantification of the emergence, and as we have shown there is a correlation between the power law exponent $\alpha$ and the memory index $q$. This, stated as a hypothesis here, may be due to the breaking of the continuous symmetry of the polymer chains and the complementarity between each pair of variables - for example temperature and pH, may mathematically take the form of two non commuting limits, with an embedded singularity as one approaches the thermodynamic limit. It may be a possibility to find a conserved Noether charge for the polymer melt in the Newtonian limit.$^{42}$ This discussion is purely intuitive and suppositions are based on our results and the theory of polymer melts, in particular HPAM needs to be investigated further from the perspective of symmetry breaking and emergence to arrive at a comprehensive solution which will determine the nature of the symmetry as well as the corresponding conserved quantity. The solution to such a problem has the potential to create a new perspective to look at fractional derivatives and thus resolve their physical inexplicability completely within a recognised framework in condensed matter theory, once again upholding generalised calculus as a quickly becoming indispensable tool in describing real-life systems.$^{19}$\\ We conclude by reiterating the most intriguing phenomenon of complementary conjugates tailoring the rheology of HPAM, which is purely intuitive from the modelling that was done. A further direction would be to attempt to find an equation of state or symmetry relationships to rigorously establish the conjecture.

\subsection*{Future Directions}

1. An open direction is to further cultivate the phenomenon of complementary conjugates by exploring symmetry relationships. It remains a possibility to find a governing equation of state, to rigorously describe what has been conjectured.

2. A continuation of this study could be made to explore the nature of the continuous symmetry that is spontaneously broken to find the exact nature of correlation between the power law exponent $\alpha$ and memory index $q$. \\

\begin{figure}[H]
\begin{center}
\includegraphics[width=9.2cm]{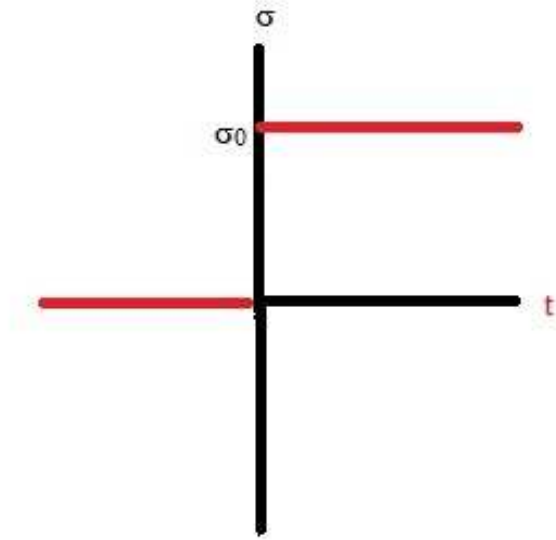}
\end{center}	
\caption{Heaviside function}	
\label{H_x_}
\end{figure}

\noindent{\textbf{Appendix}}\label{appendix}\\
The physical interpretation of a constitutive equation can be understood by performing a standard test. We consider the case for eq.(\ref{arpr4}), ie, for Maxwell Model.
To understand what this equation implies in terms of the behaviour of a tension bar under load, we subject such a bar to a two-stage standard test.\\
In first stage, known as \emph{creep phase}, we  apply at $t=0$, a constant stress, $\sigma=\sigma_0$ and observe the evolution of the strain as a function of t, ie, $\epsilon(t)$. We use Heaviside function, fig.(\ref{H_x_}) to  portray  the process of sudden loading, where,
$$\sigma= 0,~~~~~~~~ t<0,$$
$$~~=\sigma_0,~~~t>0.$$

In this case equation (\ref{arpr4}) is a differential equation for $\epsilon$ and has the solution,

\begin{equation}
\epsilon=\frac{\sigma_0}{\eta} t + c_1~~~\label{arpr5a}
\end{equation}
where $c_1$ is a constant. To find $c_1$, we impose the initial condition that, for sudden application of the stress $\sigma_0$ at $t=0,$
the initial strain produced is given by $\epsilon_0=\frac{\sigma_0}{E}$.
Evaluating for $c_1$, we get $c_1=\frac{\eta}{E}\frac{\sigma_0}{\eta}=\frac{\sigma_0}{E}.$ Thus from eq.(\ref{arpr5a})~ we obtain

\begin{equation}
\epsilon=\frac{\sigma_0}{\eta} (t+1/\lambda)~~~\label{arpr6}
\end{equation}
where, $\lambda=\frac{E}{\eta}$.\\

The second stage of the test is known as the \emph{relaxation phase}. It begins at $t=t_1$, when the strain is fixed at constant value $\epsilon=\epsilon_1$. The stress response, $\sigma(t)$ is now observed as a function of time. 
For the Maxwell model when $\epsilon=\epsilon_1,~~\dot{\epsilon}=0,$~eq.(\ref{arpr4}) is then a homogeneous differential equation for the stress $\sigma$ and has the solution

\begin{equation}
\sigma(t)=c_3 e^{-\frac{Et}{q}},~~t>t_1.~~~~~~~~~~~\label{arpr8}
\end{equation}

To find $c_3$,we reason that since the strain rate is finite everywhere, we conclude that at $t=0$, at the point of discontinuity, $\sigma(t^+)=\sigma(t^-)=\sigma_0.$~ Incorporating this into eq.(\ref{arpr8}) we evaluate $c_3$ and final expression of $\sigma(t)$ comes out to be

\begin{equation}
\sigma=\sigma_0 e^{-\frac{(t-t_1)E}{\eta}}.~~~~~~~\label{arpr9}
\end{equation}
\vspace{1em}
\noindent{\textbf{Physical Interpretation}}
\vspace{1em}

\begin{figure}[H]
\begin{center}
\includegraphics[width=9.2cm]{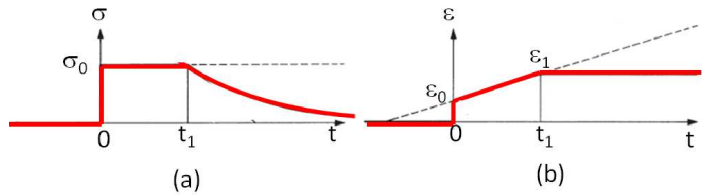}
\end{center}	
\caption{Stress and strain variation with time in the Standard Test for Maxwell Model}	
\label{test}
\end{figure}

For \textbf{Maxwell Model} it is observed that in the creep  phase (fig.\ref{test}a), the strain increases continuously as a linear function of time, starting from a non-zero value at $t=0$. The dotted lines in the $\epsilon-t$ plots in the above fig.(\ref{test}b) indicate that if the creep phase is allowed to extend beyond $t_1,$~ the strain will increase to infinity showing the fluid nature of Maxwell fluid i.e, its capability of unlimited deformation under finite stress.\\
However we also see that in eq.(\ref{arpr6}); i.e, at $t=0,~~\epsilon_0=\frac{\sigma_0}{E}$ i.e, that the response immediately after the load application is elastic with a modulus $E_0$, the initial or impact modulus.Thus the viscoelastic nature of Maxwell model is characterised.\\
In the relaxation phase (fig.\ref{test}a), the stress of the Maxwell model decreases exponentially under constant strain, implying that the material relaxes fully with stress asymptotically tending to zero.\\

There are no conflicts of interest to declare.

\end{document}